\documentclass[12pt]{article}
\usepackage{amsmath}
\usepackage{graphicx,epsfig}
\input epsf
\DeclareMathOperator{\tr}{tr}

\newcommand{\half}{\frac{1}{2}}
\newcommand{\nder}{\!\!\!\not\partial}
\begin{document}
\author{V.G. Ksenzov\footnote{State Scientific Center Institute for Theoretical and Experimental Physics, Moscow, 117218, Russia} \!\enskip and A.I. Romanov\footnote{National Research Nuclear University MEPhI, Moscow,115409, Russia}}
\title{Phase Transitions in Models with Discrete Symmetry}
\maketitle
\begin{abstract}
We investigate a class of models with a massless fermion and a self-interacting scalar field with the Yukawa interaction between these two fields. The models considered are formulated in two and four spacetime dimensions and possess a discrete symmetry. We calculate the chiral condensates are calculated in the one-loop approximation. We show that the models have a phase transitions as a function of the coupling constants.
\end{abstract}

\section{Motivation}

Dynamical breaking of a symmetry can be described by a quantity known as the order parameter. In the case of a broken chiral symmetry a commonly used order parameter is the chiral condensate which vanishes if the chiral symmetry is unbroken. The investigation of the chiral condensate plays a crucial role in attempts to describe phase transitions related to the dynamical chiral symmetry breaking.

In QCD chiral condensate is obtained by numerical studies, as a rule via lattice simulations. This approach is usually employed in studies of the chiral condensate in the presence of external factors such as temperature, chemical potentials, magnetic field, finite size effects, etc (see [1 - 3] and references therein).

Apart from numerical studies there is a lot of models employed for the investigation of the chiral condensate in various physical systems \cite{Ebert}, \cite{Bel}. In the first paper on this subject \cite{NJ} Nambu and Jona-Lasinio (NJL) analyzed a specific field model in four dimensions. Later, the chiral condensate was studied by Gross and Neveu (GN) in two-dimensional spacetime in the limit of a large number of fermion flavors $N$ \cite{GN}. These two models are similar but in contrast to the NJL model, the GN model is a renormalizable and asymptotically free theory. Due to these properties the GN model is used for qualitative modeling of QCD. The relative simplicity of both models is a consequence of the quartic fermion interaction.

In our previous paper \cite{KR} we investigated a system of a self-interacting massive scalar and a massless fermion field with the Yukawa interaction in a $(1+1)$-dimensional spacetime. In the limit of a large mass of the scalar field the model becomes equivalent to the GN model. The chiral condensate was obtained by the functional integration using the method of the saddle point. For technical reasons the contribution of the scalar field fluctuations was not taken into account. This can be done if the coupling constant of the self-interacting scalar field is much less than the Yukawa coupling constant.

Continuing our investigation of this model, we present in this paper a study of how the chiral condensate is affected by the quantum fluctuations of the scalar fields. The consideration of the quantum fluctuations allows one to find a phase transition in this model in a $(1+1)$-dimensional spacetime. Apart from that, we also investigate the chiral condensate in a $(3+1)$-dimensional spacetime. In the limit of the massless scalar field in $(3+1)$ dimensions, the chiral condensate is obtained in analytic form. We also find a phase transition in this setting. All results are obtained in the one-loop approximation framework.

\section{Chiral condensate in a $(1+1)$-dimensional spacetime}

We consider a model with the Lagrangian density given by
\begin{equation}\label{eq1}
  L=L_\mathrm{b}+L_\mathrm{f}=\half(\partial_{\mu}\phi)^2-U(\phi)+i\bar{\psi}^a\nder \psi^a-U_\mathrm{bf},
\end{equation}
where $\phi(x)$ is a real scalar field, $\psi^a$ is a massless fermion field and the index $a$ runs from 1 to $N$. The potential of the scalar field
\begin{equation}\label{eq2}
  U(\phi)=\half m^2\phi^2+V(\phi^2)
\end{equation}
includes the mass term and the self-interaction $V(\phi^2)$ of the scalar field. The potential $U_\mathrm{bf}$ is the Yukawa interaction.
$$U_\mathrm{bf}=g\phi\bar{\psi}^a\psi^a,$$
where $g$ is a coupling constant.

The Lagrangian (\ref{eq1}) is invariant under the discrete transformation
\begin{equation}\label{eq3}
  \psi^a\to \gamma_5\psi^a,\,\bar{\psi}^a\to-\bar{\psi}^a\gamma_5,\,\phi\to-\phi.
\end{equation}
The symmetry (\ref{eq3}) is broken by the chiral condensate. In this paper we extend the calculation of the chiral condensate in this model performed in ref.~\cite{KR} to include the quantum fluctuations of the scalar field.

We formally define the chiral condensate using the functional integral
\begin{equation}\label{eq4}
  \langle 0|g \bar{\psi}^a \psi ^a|0\rangle = \frac{1}{Z}\int { D \bar{\psi}^a D \psi^a D \phi g\bar{\psi}^a \psi^a e^{i \int{L(x)d^2 x}}},
\end{equation}
where $Z$ is a normalization constant. Eq.~(\ref{eq4}) can be rewritten as
$$\langle 0|g \bar{\psi}^a \psi ^a|0\rangle = \frac{1}{Z}\int  D \bar{\psi}^a D \psi^a g\bar{\psi}^a \psi^a
  \exp\left(i\int i\bar{\psi}^a \nder\psi^ad^2x\right) \times$$
\begin{equation}\label{eq5}
   \times\int D \phi \exp\left(i\int d^2x L_\mathrm{b}(x)-U_\mathrm{bf}\right).
\end{equation}
To integrate over the scalar field one decomposes $\phi=\phi_0+\varphi$, where $\phi_0$ satisfies the classical equation of motion while $\varphi$ describes small fluctuations around the classical background. In order to study the properties of the chiral condensate we choose a specific potential of the scalar field as
$$U(\phi)=\half m^2\phi^2+\frac{\lambda}{4!}\phi^4,$$
where $\lambda$ is a coupling constant. Then we take into account the quadratic terms of the scalar fluctuation $\varphi$ and integrate over them, getting
\begin{equation}\label{eq6}
  \langle 0|g \bar{\psi}^a \psi ^a|0\rangle = \frac{1}{Z}e^{iS_\mathrm{eff}^b}\cdot i\frac{\delta}{\delta\phi_0} \int  D \bar{\psi}^a D \psi^a \exp\left(i\int(i\bar{\psi}^a \nder\psi^a-g\phi_0\bar{\psi}^a\psi^a)d^2x\right),
\end{equation}
where $S_\mathrm{eff}^b=\left(\int d^2x(-U(\phi_0))\right)+i\tr\ln(\partial^2+m^2+\half\lambda\phi_0^2)$. Integrating over the fermionic field, we obtain
\begin{equation}\label{eq7}
  \langle 0|g \bar{\psi}^a \psi ^a|0\rangle =\left.\frac{Ng^2}{2\pi}\phi_0\ln\frac{g^2\phi_0^2}{\Lambda_\mathrm{f}^2}\right|_{\phi_0=\phi_m},
\end{equation}
where $\Lambda_\mathrm{f}^2$ is a fermionic ultraviolet cutoff, $\phi_m$ is a ground state value of the bosonic field which is determined by the equations
$$\left.\frac{dU_\mathrm{eff}^t(\phi_0)}{d\phi_0}\right|_{\phi=\phi_m}=0,\, \left.\frac{d^2U_\mathrm{eff}^t(\phi_0)}{d\phi_0^2}\right|_{\phi=\phi_m}>0,$$
where $U_\mathrm{eff}^t=V(\phi_0)-\frac{i}{2}\tr\ln(\partial^2+m^2+\frac{\lambda}{2}\phi_0^2)+iN\tr\ln(i\nder- g\phi_0)$, whereas $g$ and $\lambda$ are dimensional coupling constants.

It is convenient to introduce dimensionless constants $$g_0=\frac{g}{m}, \lambda_0=\frac{\lambda}{m^2}\text{ and }\Lambda^2=\frac{\Lambda_\mathrm{f}^2}{g^2}, \Lambda_s^2=\frac{\Lambda_\mathrm{b}^2}{m^2},$$
 where $\Lambda_b^2$ is a bosonic ultraviolet cutoff. Then Eq.~(\ref{eq7}) may be written as
\begin{equation}\label{eq8}
  \langle 0|g_0 \bar{\psi}^a \psi ^a|0\rangle =\frac{Ng_0^2m}{2\pi}\phi_m\ln\frac{\phi_m^2}{\Lambda^2},
\end{equation}
and the saddle point $\phi_m$ is given by the solution of the equation
\begin{equation}\label{eq9}
  1+\frac{\lambda_0}{6}\phi_m^2-\frac{\lambda_0}{8\pi}\ln\frac{1+\frac{\lambda_0}{2}\phi_m}{\Lambda_s^2}+ \frac{g_0^2N}{2\pi}\ln\frac{\phi_m^2}{\Lambda^2}=0,
\end{equation}
where all the quantities are dimensionless. The third term in this equation describes the contribution of the scalar field fluctuations. Generally speaking, Eq.~(\ref{eq9}) is resolved numerically, however, some special cases lead to simplifications that allow one to solve it analytically. An analytic solution can be easily compared with the results of our previous work \cite{KR}, where the quantum fluctuations of the scalar field were not accounted for. We consider two such possibilities.

i) when $\lambda_0=0$, then
$$\phi_m^2=\Lambda^2\exp\left(-\frac{2\pi}{g_0^2N}\right)$$
and substituting $\phi_m$ in (\ref{eq8}), we get
$$\langle 0|g_0 \bar{\psi}^a \psi ^a|0\rangle=-m\phi_m.$$
To see that this solution coincides with that of ref.~\cite{KR}, we note that $m\phi_m$ equals $\phi_m$ of that work (see eq. (15) there).

ii) when $\cfrac{\lambda_0}{2}\phi\gg 1$, then, rescaling $\cfrac{\Lambda_s^2}{\lambda_0}=\Lambda^2$ in Eq.~(\ref{eq9}), we get
$$\frac{\lambda_0}{3}\frac{2\pi\phi_m^2}{2g_0^2N-\cfrac{\lambda_0}{2}}+\ln\frac{\phi_m^2}{\mu^2}=0,$$
where
\begin{equation}\label{eq10}
  \mu^2=\Lambda^2\exp\left(-\frac{8\pi}{4g_0^2N-\lambda_0}\right).
\end{equation}
Using (\ref{eq10}) and (\ref{eq8}), we obtain
\begin{equation}\label{eq11}
  \langle 0|g_0 \bar{\psi}^a \psi ^a|0\rangle=-\frac{2Ng_0^2m}{2g_0^2N-\cfrac{\lambda_0}{2}}\left(\phi_m+ \frac{\lambda_0}{6}\phi_m^3\right).
\end{equation}
If $4Ng_0^2\gg\lambda_0$, then the value of the chiral condensate given by Eq.~(\ref{eq11}) equals that of Eq. (13) in ref.~\cite{KR}.

The value of $\mu^2$ is renormalization invariant. Indeed, the $\beta$-functions in the one-loop approximation are
\begin{equation}\label{eq12}
  \beta(g_{0})=\Lambda\frac{dg_{0R}}{d\Lambda}=\frac{g_0}{4\pi}\left(\frac{\lambda_0}{2}-2g_0^2N\right),\, \beta(\lambda_0)=\Lambda\frac{d\lambda_{0R}}{d\Lambda}=\frac{\lambda_0}{2\pi} \left(\frac{\lambda_0}{2}-2g_0^2N\right).
\end{equation}
Using (\ref{eq12}) one can verify that $\mu^2$ does not depend on the choice of the renormalization point, i.e.
$$\Lambda\frac{d\mu^2}{d\Lambda}=0.$$
It is worth to note that in a $(1+1)$-dimensional spacetime there is only one $Z$-factor that appears in the renormalization of the scalar field mass, therefore
\begin{equation}\label{eq13}
  \frac{g_{0R}^2(\Lambda)}{\lambda_{0R}(\Lambda)}=\frac{g_0^2}{\lambda_0},
\end{equation}
where $g_0^2$ and $\lambda_0$ are the bare coupling constants, and $g_{0R}^2$, $\lambda_{0R}$ are the renormalized coupling constants. The relation in Eq. (13) can be obtained directly from the $\beta$-functions of Eq.~(\ref{eq12}).

Using Eq.~(\ref{eq13}) $\mu^2$ is rewritten as
\begin{equation}\label{eq14}
  \mu^2=\Lambda^2\exp\left(-\frac{4\pi}{2Ng_{0R}^2}\left(1-\frac{\lambda_0}{4Ng_0^2}\right)^{-1}\right).
\end{equation}
One can see that $\mu^2$ and the $\beta$-functions tend to zero in the limit $\lambda_0\to 4Ng_0^2$. On the other hand, if the $\beta$-function equals to zero then in the model there is a phase transition as shown in ref.~\cite{Pol}. Therefore one can conclude that if $\cfrac{\lambda_0}{4Ng_0^2}$ is $O(1)$ then there is a phase transition in the model. Although $\mu^2$ is not a solution of the Eq.~(\ref{eq9}) or Eq.~(\ref{eq10}), $\mu^2$ also tends to zero when the $\beta$-functions tend to zero therefore $\mu^2$ may be considered as a qualitative approximation of $\phi_m$. The value of $\mu^2$ becomes a good enough approximation of $\phi_m$ provided $\cfrac{\lambda_0}{4Ng_0^2}\ll 1$.

The results of our numerical studies of the chiral condensate are presented in Fig. 1. For the convenience the value $R=-\langle 0|g_0 \bar{\psi}^a \psi ^a|0\rangle$ is shown in this figure. We plot the value $R$ as a function of the Yukawa coupling constant $g_0$. The other parameters entering Eq.~(\ref{eq9}) take the values $N=5, \lambda_0=0.5, \Lambda^2=20, m=0.2$. Since the chiral condensate falls sharply when $\cfrac{\lambda_0}{4Ng_0^2}$ tends to 1, then $R$ is presented in different scales in Fig. 1 (a,b). The solid line corresponds to $\phi_m$ obtained from the solution of Eq.~(\ref{eq9}). The dotted line corresponds to $\phi_m$ obtained without the contribution of the scalar field fluctuations. The dash-dotted line is obtained for $\phi_m=\mu$. We see that the chiral condensate equals to zero and the chiral symmetry is unbroken provided $\cfrac{\lambda_0}{4Ng_0^2}$ is $O(1)$. The chiral condensate appears when this condition is violated and increases with $g_0^2$. Hence there is a phase transition in this model. With increasing $g_0^2$ the quantity $\mu^2$ becomes a good approximation of $\phi_m$ according to the corresponding condition $\cfrac{\lambda_0}{4Ng_0^2}\ll 1$.

\begin{figure}
\centering
\includegraphics[width=\linewidth]{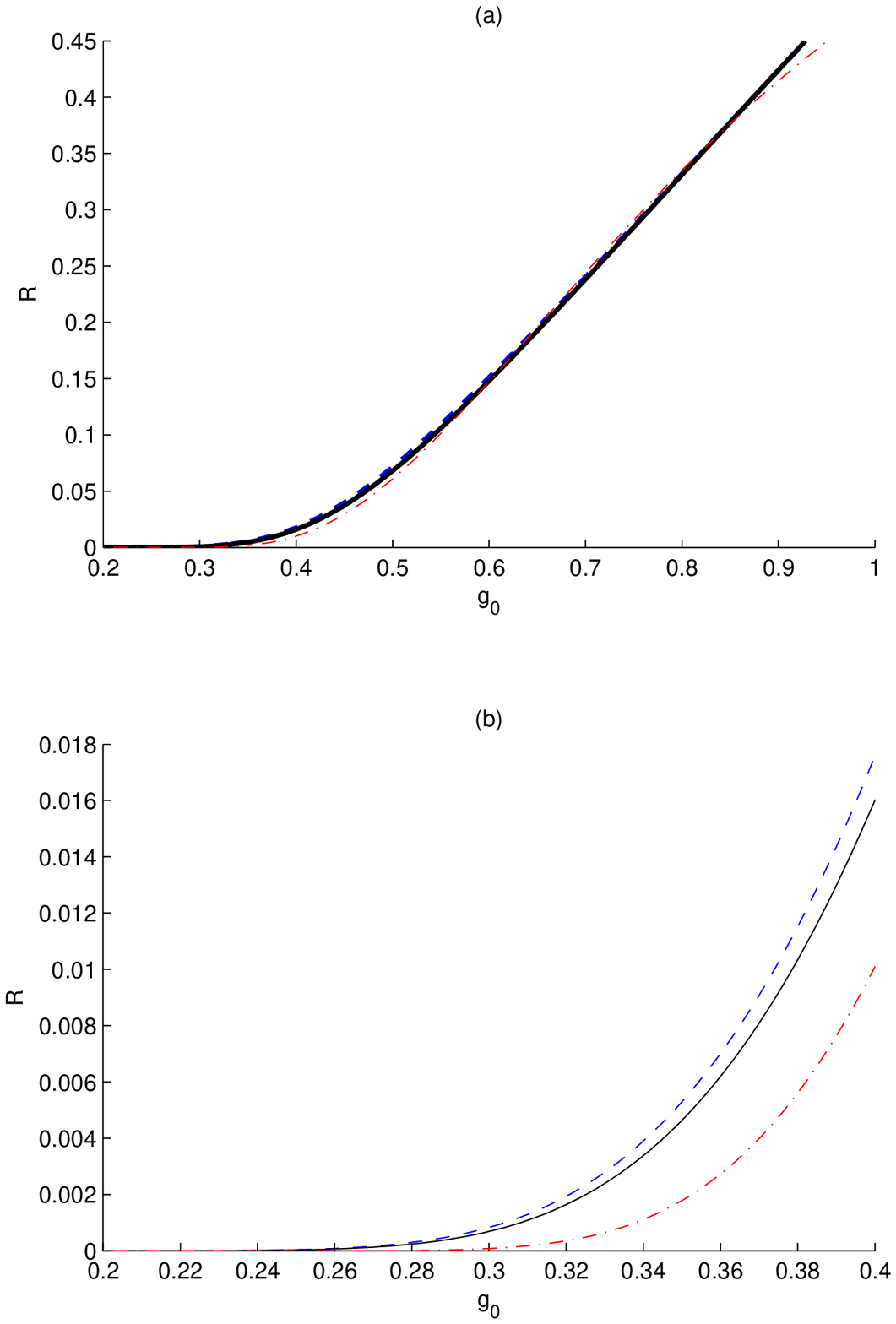}
\caption{The value $R=-\langle 0|g_0 \bar{\psi}^a \psi ^a|0\rangle$ as a function of the Yukawa coupling constant $g_0$. The values of parameters are $N=5, \lambda_0=0.5, \Lambda^2=20, m=0.2$. The solid line shows the chiral condensate where $\phi_m$ is an exact solution of Eq.~(\ref{eq9}). The dotted line is $R$ obtained without the contribution of the scalar field fluctuations. The dash-dotted line is $R$ obtained for $\phi_m=\mu$.}
\label{Fig1}
\end{figure}

\section{Chiral condensate in a $(3+1)$-dimensional spacetime}

The method used above to calculate the chiral condensate is applicable in any dimension of space. Therefore calculating $U_\mathrm{eff}$ we get
$$U_\mathrm{eff}(\phi)=\half m^2\phi^2+\frac{\lambda}{24}\phi^4+\frac{\left(m^2+\half \lambda\phi^2\right)^2} {64\pi^2}\left(\ln\frac{m^2+\half\lambda\phi^2}{\Lambda_\mathrm{b}^2}-\half\right)-$$
\begin{equation}\label{eq15}
-\frac{Ng^4}{16\pi^2}\phi^4 \left(\ln\frac{g^2\phi^2}{\Lambda_\mathrm{f}^2}-\half\right).
\end{equation}
The quadratically divergent terms which appear in $U_\mathrm{eff}(\phi)$ are eliminated by the mass renormalization.

Further we consider the case of a massless scalar field $m^2=0$. Rescaling the ultraviolet cutoff
$$\frac{2\Lambda_\mathrm{b}^2}{\lambda}=\frac{\Lambda_\mathrm{f}^2}{g^2}=\Lambda^2,$$
we rewrite the effective potential (\ref{eq15}) as
\begin{equation}\label{eq16}
  U_\mathrm{eff}(\phi)=\frac{\lambda\phi^4}{24}+\frac{\lambda^2\phi^4}{256\pi^2}\left(\ln\frac{\phi^2} {\Lambda^2}-\half\right)-\frac{Ng^4\phi^4}{16\pi^2}\left(\ln\frac{\phi^2}{\Lambda^2}-\half\right)
\end{equation}
or
\begin{equation}\label{eq17}
  U_\mathrm{eff}(\phi)=\frac{\lambda^2-16g^4N}{256\pi^2}\phi^4\left(\ln\frac{\phi^2}{\mu^2} -\half\right),
\end{equation}
where
\begin{equation}\label{eq18}
  \mu^2=\Lambda^2\exp\left(-\frac{32\pi^2\lambda}{3(\lambda^2-16Ng^4)}\right).
\end{equation}
Note that $\mu^2$ is not a renormalization-invariant quantity. It is possible to check that by direct calculation, but it can be done in a simpler way. Namely, we consider the following limiting cases

i) $\lambda=0$, $\mu^2=\Lambda^2$,

ii) $g^2=0$, $\mu^2=\Lambda^2\exp\left(-\cfrac{32\pi^2}{3\lambda}\right)$ is not a renormalization-invariant quantity, we recall that $\beta=\cfrac{3\lambda^2}{32\pi^2}$.

In four dimensions the chiral condensate is defined as
$$\langle 0|g \bar{\psi}^a \psi ^a|0\rangle =\left.\frac{Ng^4\phi^3}{4\pi^2}\ln\frac{\phi^2}{\Lambda^2}\right|_{\phi=\phi_m},$$
where $\phi_m$ is determined by the equations
\begin{equation}\label{eq19}
  \left.\frac{dU_\mathrm{eff}(\phi_0)}{d\phi_0}\right|_{\phi=\phi_m}=0,\, \left.\frac{d^2U_\mathrm{eff}(\phi_0)}{d\phi_0^2}\right|_{\phi=\phi_m}>0.
\end{equation}
The effective potential $U_{eff}$ has a minimum at $\phi_m=\mu^2$ and $\lambda^2>16Ng^4$. In this case the chiral condensate is given by
\begin{equation}\label{eq20}
  \langle 0|g \bar{\psi}^a \psi ^a|0\rangle =-\frac{8Ng^4\lambda}{3}\frac{\mu^3}{\lambda^2-16Ng^4}.
\end{equation}
In the limit $\lambda^2\to 16Ng^4$, $\mu^2$ tends to zero and consequently $\langle 0|g \bar{\psi}^a \psi ^a|0\rangle\to 0$. If $16Ng^4>\lambda^2$ then $U_\mathrm{eff}$ has a maximum, the chiral condensate does not exist, and the chiral symmetry is unbroken.

\section{Conclusions}

In this paper we studied the dynamical symmetry breaking in models consisting of a fermion field and a self-interacting scalar field with the Yukawa interaction between the two fields in the one-loop approximation. We believe that the dynamical symmetry breaking as exhibited by these models can help to analyze similar phenomena in more realistic models. The models are considered in two and four spacetime dimensions. We showed that each models has two phases, one with the preserved symmetry and the other with the broken symmetry. We conclude that both models have a phase transitions as a function of the coupling constants. In four spacetime dimensions, we obtain an analytic expression for the chiral condensate in the limit of a massless scalar field. The two-dimensional case was investigated numerically by solving of the algebraic equation for the scalar field.

\begin{center}
Acknowledgments
\end{center}

We are grateful to O.V. Kancheli and V.A. Lensky for useful discussions.

\end{document}